\documentclass[12pt,epsfig,preprint]{aastex}

\lefthead{B. Paul et al.}
\righthead{SMC~X-1, LMC~X-4}

\begin{document}

\title{
Nature of the Soft Spectral Component in the X-ray Pulsars
SMC~X-1 and LMC~X-4}

\author{B. Paul\altaffilmark{1,2}, F. Nagase\altaffilmark{1},
T. Endo\altaffilmark{1}, T. Dotani\altaffilmark{1},
J. Yokogawa\altaffilmark{3}, and  M. Nishiuchi\altaffilmark{3}}

\altaffiltext{1}{The Institute of Space and Astronautical Science,
3-1-1 Yoshinodai, Sagamihara, Kanagawa 229-8510, Japan, nagase@astro.isas.ac.jp,
 dotani@astro.isas.ac.jp, etakao@isl.melco.co.jp}

\altaffiltext{2}{Tata Institute of Fundamental Research,
Homi Bhabha road, Mumbai, 400\,005, India bpaul@tifr.res.in}
\altaffiltext{3}{
Department of Physics, Graduate School of Science, Kyoto University,
Sakyo-ku, Kyoto, 606-8502, Japan jun@cr.scphys.kyoto-u.ac.jp, nishiuchi@apr.jaeri.go.jp}

\authoraddr{Department of Astronomy and Astrophysics,
Tata Institute of Fundamental Research, Homi Bhabha road, Mumbai,
400\,005, India}

\date{}

\begin{abstract}

We present here the results of an investigation of the pulse averaged
and pulse phase resolved energy spectra of two high luminosity accretion
powered X-ray pulsars SMC~X-1 and LMC~X-4 made with ASCA. The phase
averaged energy spectra definitely show the presence of a soft excess
in both the sources. If the soft excess is modeled as a separate
black-body or thermal bremsstrahlung type component, pulse phase resolved
spectroscopy of SMC~X-1 shows that the soft component also has a
pulsating nature. Same may be true for LMC~X-4, though a very
small pulse fraction limits the statistical significance.
The pulsating soft component is found to have a nearly sinusoidal profile,
dissimilar to the complex profile seen at higher energies, which can
be an effect of smearing.
Due to very high luminosity of these sources, the size of the emission
zone required for the soft component is large (radius $\sim$300--400 km).
We show that the pulsating nature of the soft component is difficult to
explain if a thermal origin is assumed for it. We further investigated
with alternate models, like inversely broken power-law or two different
power-law components and found that these models can also be used to
explain the excess at low energy. A soft power-law component may be a
common feature of the accreting X-ray pulsars, which is difficult to
detect because most of the HMXB pulsars are in the Galactic plane and
experience large interstellar absorption. In LMC X-4, we have also
measured two additonal mid-eclipse times, which confirm the known
orbital decay.

\end{abstract}


\section{Introduction}

The X-ray continuum spectra of accreting pulsars are often described as a
broken power-law or a power-law with exponential cutoff. The break
in the spectrum is in the range of 10--20 keV and power-law photon index
below the break energy is in the range of 0--1 (White, Nagase \& Parmar
1995). Some binary X-ray pulsars which are away from the Galactic plane and
therefore experience less interstellar absorption, show the presence of
a soft component in the spectrum which is often modeled as a black-body
and/or thermal bremsstrahlung emission (SMC~X-1: Marshall, White, \& Becker
1983; Woo et al. 1995; Wojdowski et al. 1998; LMC~X-4: Dennerl 1989;
Woo et al. 1996; RX~J0059.2--7138: Kohno, Yokogawa,
\& Koyama 2000; 4U~1626--67: Orlandini et al. 1998; Her~X-l: McCray et al.
1982, Dal Fiume et al. 1998; Oosterbroek et al. 1997, 2000; Endo, Nagase
\& Mihara 2000) or an inversely broken power-law (XTE~J0111.2--7317:
Yokogawa et al. 2000b). The soft component in 4U~1626--67, when modeled
as a black-body emission requires the size of emission region to be
comparable to that of the neutron star because the intrinsic luminosity
of this source is of the order of 10$^{35}$ erg s$^{-1}$ (Orlandini et al.
1998). On the other hand, the soft component in Her~X-1 can be modeled as a
black-body which is reprocessed emission from the innermost part of the
accretion disk (Endo et al. 2000). However, for the bright pulsars in the
Magellanic Clouds for which the distance is of the order of 50--60 kpc
and the luminosity is close to the Eddington limit, the luminosity of
the soft excess is about an order of magnitude larger compared to the
same of Her~X-1. Therefore, a pulsating nature of the soft
component that has been observed in some high luminosity X-ray pulsars
(LMC~X-4 : Woo et al. 1996 and XTE~J0111.2--7317 : Yokogawa et al.
2000b) needs to be probed with greater detail.

To investigate the pulse phase dependence of the soft component of
accreting X-ray pulsars in detail,
we have chosen two luminous X-ray pulsars SMC~X-1 and LMC~X-4, which are in
the Magellanic Clouds and suffer less interstellar absorption.
SMC~X-1 and LMC~X-4 are two bright, eclipsing, accreting, binary X-ray
pulsars with spin periods of $\sim$0.7 and $\sim$13.5 s and binary periods
of $\sim$3.9 and $\sim$1.4 day respectively. The companion of SMC~X-1
is a B0 supergiant while the companion star of LMC~X-4 is of type O7III-V.
The binary orbits of both the systems are nearly circular. The orbital
period of the two binaries are found to decay with time scale of
3 $\times$ 10$^5$ yr in SMC~X-1 (Wojdowski et al. 1998), and
10$^6$ yr in LMC~X-4 (Levine, Rappaport \& Zojcheski 2000).
Another striking similarity between these two sources is a
long-period of 50--60 day and 30.5 day respectively, that is known to be
quasi-stable in SMC~X-1 (Wojdowski et al.  1998) and stable in LMC~X-4
(Lang et al. 1981).
The long-period is believed to
be a result of (quasi) periodic obscuration of the neutron star by a
precessing accretion disk, similar to that in Her~X-1. Broad band
energy-spectra (0.2--37 keV) of these two sources were studied by performing
a combined fit to the
observations made with ROSAT and GINGA (Woo et al. 1995; 1996). In
addition to a
cutoff power-law type component in SMC~X-1 (photon index of $\sim$0.93,
$E_{\rm C}$ = 5.6 keV, $E_{\rm F}$ = 15 keV)
and
power-law type component in LMC~X-4 (photon index of $\sim$0.67)
broad iron emission lines and
soft excess were detected. The soft component was modeled
as a single black-body component ($kT_{\rm BB}$ = 0.16 keV) in SMC~X-1 and
as a sum of a low
temperature black-body emission ($kT_{\rm BB}$ = 0.03 keV) and a thermal
bremsstrahlung emission ($kT_{\rm TB}$ = 0.35 keV) in LMC~X-4.
Beppo-SAX observations also showed presence of similar soft component
(La Barbera et al. 2001).
From HEAO-1 observations, the soft X-rays from SMC X-1 were found
to be nonpulsating (Marshal et al. 1983). However, the ROSAT and ASCA
observations detected clear pulsations with a pulse profile different
from the hard component (Wojdowski et al. 1998). Pulse
phase resolved spectroscopy of combined ROSAT and GINGA data of LMC~X-4
revealed modulation of the thermal bremsstrahlung and iron line emission
components with pulse phase (Woo et al. 1996).

We note that the combined fit of the ROSAT and GINGA spectra of
these two sources were performed on non-simultaneous observations and the
intensity and spectral shape of these sources are known to be variable.
We have carried out pulse phase averaged and pulse phase resolved spectral
studies in the 0.5--10.0 keV band. In this paper we present the results of
our investigation to the nature of the soft excess through pulse profiles in
different energy bands and variations of the different spectral components
with pulse phase.

\section{Observations}

The observations of SMC~X-1 and LMC~X-4, presented here were made with
the Advanced Satellite for Cosmology and Astrophysics (ASCA). ASCA is
equipped with two Solid-state Imaging Spectrometers (SIS) and Gas Imaging
Spectrometers (GIS), each at the focal plane of four identical mirrors of
typical photon collecting area $\sim$150 cm$^2$ at 6 keV. The energy resolution
is 130 and 500 eV (FWHM) at 6 keV for the SIS and GIS detectors
respectively. For more details about ASCA refer to Tanaka, Inoue \& Holt
(1994).

SMC~X-1 was observed twice with ASCA during 1993 April 16--27, and 1995
October 18--19, and it was at the edge of the GIS field of view in an
observation of XTE~0111.2--7317 made during 1998 November 18--19
(Yokogawa et al. 2000a; 2000b). The first observation was in the high
state of its long-period intensity variation of 50--60 day, while the
second one was in the low state. Results from timing, and some spectral
analysis of these observations were presented by Wojdowski et al. (1998)
and Yokogawa et al. (2000a). For the present study, we have selected the
first observation made in 1993 which has the best photon statistics. In
this observation, the GIS detectors were operated in normal PH mode with
time resolution of 1.953 ms and 15.625 ms for high and medium bit-rate,
respectively, while the SIS observations were
made with one of the CCD chips, with time resolution of 4 s.
Out of eclipse data obtained during the first 60 ks of the observation
resulted into 20.5 ks and 18 ks of useful data with each of the
GIS and SIS detectors respectively. The GIS and
SIS spectra have been used for phase averaged spectroscopy while for
phase resolved studies only data from GIS could be used. Wojdowski et al.
(1998) determined the pulse periods and other orbital parameters of
SMC~X-1 from the ASCA observations.

LMC~X-4 was observed with ASCA on three occasions, 1994 April
26--27, 1995 November 24--25, and 1996 May 24--25, all during the high state
of its 30.5 day long-period. The second observation was carried out mainly
during the eclipse. Of the remaining two, we have selected the third
observation for the present study, because there were some dips in the
first observation with associated spectral changes which makes the 
analysis more complicated. For analysis of the orbital period decay in
LMC X-4, however, both the 1994 and 1996 observations were used. Some
preliminary results from some of these
observations were reported by Vrtilek et al. (1997). In the 1996
observation, the GIS detectors were operated in normal PH mode in which
the time resolution is 62.5 ms and 500 ms at high and medium bit rates
respectively. Total exposure of 36 ks was obtained with the GIS, of
which 31 ks of observation is outside the eclipse. Observations with
the SIS detectors were made with one of the CCD chips. SIS0 was operated
in FAST mode with time resolution of 15.625 ms while SIS1 was operated
in BRIGHT (imaging) mode which has a time resolution of 4 s. The average
total exposure and out of eclipse exposure with the two SIS detectors were
33.5 ks and 28.5 ks respectively. The GIS and SIS1 data outside the
eclipse have been used for phase averaged spectroscopy while GIS data
obtained only in the high bit rate and SIS0 FAST mode data have been used
for timing and pulse phase resolved spectroscopy.
The pulse periods of LMC~X-4 obtained from these ASCA observations
have been reported by Vrtilek et al. (1997).

\section{Analysis and results}

We used the standard data selection criteria of the ASCA guest observer
facility. Data from the hot and flickering pixels of the SIS detectors
were removed. Charged particle events were removed from the GIS data based on
the rise time discrimination method. The count rates and spectra of the source
were obtained from circular regions of radius 6$\arcmin$ for the GIS detectors
and 4$\arcmin$ for the SIS detectors. For SIS, the background spectra were
extracted from the whole chip excluding a circular region around the source
and for GIS it was collected from regions diametrically opposite to the
source location in the field of view.

\subsection{SMC~X-1}
\subsubsection{Pulse profile}

This observation was made in 1993 April 16--27. Near the end of this
observation, the source went into an eclipse and the count rate shows
a gradual decrease. The light curve obtained from this observation is
shown in Figure 1 for different energy bands.  In the present work, we
have used data only from the first 60 ks when the source was out of
eclipse and the count rate was high.

To create the pulse profiles, and to carry out pulse phase resolved
spectroscopy, the photon arrival times were reassigned after solar
system barycenter correction and correction for the arrival time
delays due to orbital motion. The orbital parameters and the pulse
period were taken from Wojdowski et al. (1998). The pulse profile,
shown in Figure 2,
is single peaked at energies $<$ 1.0 keV, with an additional small peak
at higher energies. The pulse phases are identical in different energy
bands. The pulse fraction is somewhat larger at higher energies.

\subsubsection{Phase averaged spectrum}

After appropriate background subtraction, pulse phase averaged spectra
from the pairs of GIS and SIS detectors were combined and were fitted
simultaneously.
The relative normalisation of the SIS detectors was allowed to vary
and all the other spectral
parameters were tied to be the same for SIS and GIS. The energy ranges
chosen for spectral fitting are 0.55--10.0 keV for the SIS and 0.7--10.0
keV for the GIS. All the spectral analysis reported in this paper were
carried out using the spectral analysis package XSPEC.

The spectra, in addition to a hard power-law component with photon index
$\sim$0.9, also showed iron line and soft excess. We found that to
explain the soft excess, a second component is needed.
The soft excess can be fitted to either a (a) black-body, (b)
bremsstrahlung, (c) soft power-law, or (d) inversely broken power-law.
We found that a partial covering model does not fit the
spectrum well.
In SMC~X-1, the hard power-law component is known to have an exponential
cutoff at 6 keV with a $e$-folding energy of $\sim$25 keV (Woo et al. 1996).
This was also found in the ASCA data and a high energy exponential
cutoff component was included.
Analytic form of the different models
used are as follows:
\begin{eqnarray*}
{\rm Model~I~:}~f(E) &=& e^{-\sigma(E) N_{\rm H}}
         \left(f_{\rm PF}(E) + f_{\rm PL}(E)  f_{\rm cut}(E)
	+ f_{\rm Fe}(E) \right),\\
{\rm Model~II~:}~f(E) &=& e^{-\sigma(E) N_{\rm H}}
         \left(f_{\rm TB}(E) + f_{\rm PL}(E)  f_{\rm cut}(E)
	+ f_{\rm Fe}(E) \right),\\
{\rm Model~III~:}~f(E) &=& e^{-\sigma(E) N_{\rm H}}
         \left(~~~~~~f_{\rm PL_{\rm BKN}}(E)  f_{\rm cut}(E)~~~~
	+ f_{\rm Fe}(E) \right),\\
{\rm Model~IV~:}~f(E) &=& e^{-\sigma(E) N_{\rm H}}
         \left(f_{\rm PL_{\rm S}}(E) + f_{\rm PL}(E)  f_{\rm cut}(E)
	+ f_{\rm Fe}(E) \right),\\
f_{\rm PF}(E) &=& {{I_{\rm PF} \left({{E}\over{kT_{\rm PF}}} \right)^2(e-1)}
\over{{{e^{{E}\over{kT_{\rm PF}}}} -1 }}},~~~~
f_{\rm TB}(E) = I_{\rm TB}{{G(E,kT_{\rm TB})}\over{E}}e^{-{{E}
\over{kT_{\rm TB}}}},\\
f_{\rm PL}(E) &=& I_{\rm PL}E^{-\Gamma_1},~~~~
f_{\rm PL_{\rm S}}(E) = I_{\rm PL_{\rm S}}E^{-\Gamma_2},\\
f_{\rm PL_{\rm BKN}}(E) &=& \left\{
                 \begin{array}{@{\,}l@{\,}}
                 I_{\rm P} E^{-\Gamma_2} \ (E < E_{\rm b})   \\
                 I_{\rm P} E_{\rm b}^{(\Gamma_1 - \Gamma_2)}E^{-\Gamma_1} \
                 (E \geq E_{\rm b};\ \Gamma_1 < \Gamma_2)   \\
                 \end{array}
                 \right\},\\
f_{\rm cut}(E) &=& \left\{
                 \begin{array}{@{\,}l@{\,}}
                 1 \ (E < E_{\rm C})   \\
                 e^{-{{E - E_{\rm C}}\over{E_{\rm F}}}} \
                 (E \geq E_{\rm C})\\
                 \end{array}
                 \right\},\\
f_{\rm Fe}(E) &=& \frac{I_{\rm Fe}}{\sqrt{2 \pi \sigma_{\rm Fe}^2}}
                  \exp{[-\frac{(E-E_{\rm Fe})^2}{2\sigma_{\rm Fe}^2}]},\\
\end{eqnarray*}                                                                 
where $E$ is the incident photon energy, $\sigma(E)$ is the photo-electric
absorption cross-section (Morrison \& McCammon 1983),
$G(E, kT_{\rm TB})$ is the Gaunt factor at energy $E$ for a plasma of
temperature $kT_{\rm TB}$ keV.
$E_{\rm C}$ and $E_{\rm F}$ are the cutoff energy and the $e$-folding
energy respectively.

The phase averaged spectrum could be fitted well
with all the four models and the parameters obtained for each model are
given in Table 1. The photon index ($\Gamma_1$) of the hard power-law component
is identical in all the four models, $\sim$0.9. The cutoff energy is found
to be 5.5 keV and the e-folding energy above the cut-off is 25 keV.
In SMC~X-1, a black-body type model for the soft component requires a
temperature of 0.18 keV and for a distance of 65 kpc of the SMC, it
should have a radius of $\sim$400 km. The SIS and GIS count rate spectra of
SMC~X-1 are shown in Figure 3 along with the components of model-I and
the residuals. A thermal bremsstrahlung type emission for the soft
excess requires a temperature of 0.33 keV and emission measure of
$7 \times 10^{61}$ cm$^{-3}$. In case of an additional power-law,
it has a photon
index of 4.8, whereas in inversely broken power-law model, the soft photon
index is $\sim$1.9 and the break energy is $\sim$1.7 keV. The iron
K$_\alpha$ line has an equivalent width of 120 $\pm$ 30 eV and it
is found to be broad (gaussian $\sigma$ = 0.5 keV).

\subsubsection{Pulse phase resolved spectrum}

To extract the phase resolved spectra, phase filtering was applied withtin
the FTOOLS task XSELECT. In this method, the photon arrival time or the
center time of each data bin is used to determine which phase bin the photon
belongs to.
In order to search for variability of the soft component, the phase resolved
spectra were fitted with the models I and II which have a thermal
component for the soft excess. A pulsating nature of the soft component
has been detected in both the models with identical amplitude. Here we
discuss only the results obtained with the first model.
At first we created four phase resolved spectra consisting the main pulse,
sub pulse, and two minima of the pulse profile as shown in the bottom
panel of Figure 2. While fitting these four spectra, all the spectral
parameters were allowed to vary. The results of this analysis are shown
in Figure 4 and Table 2. While the normalisation of the power-law and
black-body components are found to have considerable variation (Table 2),
variations of the black-body temperature and photon index are within
measurement uncertainities. 
Since we have used GIS data in the energy range 0.7-10.0 keV,
the column density values are not well constrained to see any measurable
variation (Table 2).
We, therefore, fixed most of the spectral parameters to their phase
averaged value during the later part of the phase resolved spectroscopy,
when we have more phase bins and poorer statistics in each phase bin.

Further, 16 pulse phase resolved spectra were
obtained only from the GIS data. The
variations of the power-law flux, black-body flux, and the total
flux with the pulse phase are shown in Figure 5.
It is clearly seen that the strength of the power-law component has a
profile similar to the pulse profile. The black-body component shows a
nearly sinusoidal intensity variation with the pulse phase.
There is also phase difference between the soft and hard pulse profiles.
A constant fit to the phase resolved black-body flux gives a reduced
$\chi^{2}$ of 7.2 and a sinusoidal fit improves the reduced $\chi^{2}$
to 4.0. While a sinusoidal does not represent the variation of the
black-body flux very well, poorness of the first fit certainly shows
a varying nature of the black-body flux.

Assuming that the pulse profiles at different narrow energy bands are
made of two components, one nearly sinusoidal and the other complex
(shown in the top and middle panels of Figure 5), we have estimated the
fraction of sinusoidal and complex components in the SMC~X-1 pulse
profile at different energies.
The narrow band pulse profiles were fitted to a composite of two
templates (made from the top and middle panels of Figure 5) and
the fractions of the two components that give minimum $\chi^{2}$
was found out.
In Figure 6, we compare this with the
ratio of the black-body to total flux derived from the best fit spectral
model. A resonable agreement between the two provides a consistency
for the hypothesis that there are two spectral components with different
pulse profiles.

\subsection{LMC~X-4}
\subsubsection{Orbital period decay}

The orbital period of  LMC~X-4 is known to have a decay time scale of
10$^6$ yr. The decay rate of the binary orbit of LMC~X-4
has been established with high confidence by combining the earlier
known mid-eclipse times with a recent precise measurement with the
RXTE (Levine et al. 2000). From the ASCA observations of LMC~X-4, we
have determined two more mid-eclipse times. Among the four orbital
parameters, orbital period, semi-amplitude of the arrival time delay,
eccentricity and mid-eclipse time, the later is likely to have maximum
variation between observation to observation. This also allows precise
measurement of the period change. A small pulse fraction and low count
rate of LMC~X-4 with ASCA, does not allow us to directly apply the pulse
arrival time analysis. We, therefore adopted a different approach to
determine the mid-eclipse time.

After the barycentric
corrections, the light curves were corrected for delays due to the orbital
motion with the known semi-amplitude, orbital period and zero eccentricity
of the system
for various trial mid-eclipse times around its extrapolated value. The
epoch-folding technique was applied on each of these light curves 
(with time resolution of 62.5 ms) and
the maximum $\chi^2$ and the pulse period corresponding
to each trial mid-eclipse time was determined. The distribution of
maximum $\chi^2$ against the trial mid-eclipse times are shown in
Figure 7 for the two ASCA observations in 1994 and 1996.
The maximum $\chi^2$ distribution has a gaussian profile around its
expected value, the center of which gives the correct mid-eclipse time.
The two mid eclipse times obtained are MJD 49468.6859 $\pm$ 0.0054 and
MJD 50227.8069 $\pm$ 0.0016 and the corresponding pulse periods are
13.5075 $\pm$ 0.0002 s and 13.5088 $\pm$ 0.0001 s respectively.
In absence of a clear method of error
determination, we have taken the width of the gaussians as the
error of the mid-eclipse time, which possibly is an overestiamtion.

The mid-eclipse times determined here are consistent with the orbital
evolution history of LMC~X-4
with evolution time scale of 10$^6$ yr (Levine et al. 2000).
The pulse profiles obtained from the GIS detectors are shown in Figure 8
for two different and summed energy bands. The profile is a single peaked
sinusoidal in the low energy band of 0.5--1.5 keV and is complex with
multiple peaks in the higher energy band of 1.5--10.0 keV.
For pulse phase resolved spectroscopy, orbital motion of the neutron
star was corrected in the light curve using the known semi-amplitude,
eccentricity (Levine et al.  2000) and mid-eclipse time derived from
the ASCA data.

\subsubsection{Phase averaged spectrum}

The four spectral models used for fitting the SMC~X-1 spectrum were
modified slightly for the LMC~X-4 spectrum. The high energy cutoff
component was removed. Residuals of the best fit LMC~X-4 spectra with
the four above mentioned models showed some line like features around
1 and 2 keV.
Therefore, while fitting the LMC~X-4 spectrum
we included two gaussian components.


Similar to SMC~X-1, the phase
averaged spectrum could be fitted well with all the four models described
above and the parameter values for the best fit obtained for each model
are given in Table 3. The photon index ($\Gamma_1$) of the hard power-law
component is identical in all the four models, $\sim$0.70. A black-body
for the soft component requires a temperature of 0.17 keV and for a
distance of 55 kpc of the LMC, if the black-body emission region is
assumed to be spherical, it should have a radius of $\sim$300 km. In
Figure 9, the SIS and GIS count rate spectra are shown along with the
components of model-I in the upper panel, while the residuals to the best
fit model are shown in the bottom panel. A thermal bremsstrahlung emission
with a temperature of 0.5 keV and emission measure of 5$\times$ $10^{60}$
cm$^{-3}$
also fits the soft excess well.
If the soft excess is modeled in the form of an
additional power-law, it has a photon index of 2.9, whereas an inversely
broken power law type spectrum requires a soft photon index of $\sim$1.9
below a break energy of 1.7 keV.
Compared to the broken power-law model, the two power-law model, which
has a cross-over energy of 1.5 keV, gives a 
steeper soft power-law.
In addition, there are line-like features
at $\sim$1.0 and $\sim$1.9 keV which we tentatively identify as iron
L$_\alpha$ and ${\rm Si_{XII}}$. The width of these two low energy lines were
fixed at 0.1 keV during the spectral fit. The iron K$_\alpha$ line is
centered at 6.4 keV and is broad (gaussian $\sigma$ = 0.65 keV), with an
equivalent width of 140 $\pm$ 40 eV. The
L/K line ratio is model dependent, and is in the range of 2--5. 

Woo et al. (1996) and La Barbera et al. (2001) fitted the soft excess
in LMC~X-4 as thermal bremsstrahlung emission. The temperature and
flux of the thermal bremsstrahlung component obtained in Model II is
similar to the their measurements, which were also carried out in
high state of LMC~X-4. Same is true about the flux and equivalent
width of the iron emission line.
La Barbera et al. (2001)
found evidence of two distinct iron emission lines from the
Beppo-SAX observation and the 6.1 keV line remains unexplained.
However, an independent analysis of the same
data (Oosterbroek et al. 2002, private communication) was found to favour
a model consisting of a hard power-law, a black body type soft component,
an emission line at 0.9 keV and a single iron emission line.
The residuals of the best fitted model spectrum from the ROSAT and
GINGA data (Figure 5 of Woo et al. 1996) also show the possibility of
two weak emission lines at $\sim$ 1 keV and $\sim$ 2 keV.

\subsubsection{Pulse phase resolved spectrum}

Similar to the case with SMC~X-1,
only four parameters related to the shape and strength of the hard
power-law and the soft excess were allowed to vary while all the other
spectral parameters were fixed to their phase averaged value.

After the barycentric and orbital motion corrections of the photon arrival
times, spectra were extracted from 16 uniformly divided pulse phases. 
All but the photon index and normalisation
of the hard power-law component and temperature and normalisation of
the soft component were fixed to their best fit values obtained from the
pulse averaged spectrum. Modulations of flux from the hard and the soft
components are shown in Figure 10 along with the total flux. The phase
resolved spectra show that the power-law component follows the
high-energy ($>$ 1.5 keV) pulse profile with three peaks and the soft
component (either black-body or bremsstrahlung)
is consistent to a single peaked sinusoidal modulation
similar to the pulse profile in the soft band
($<$ 1.5 keV).
For the SIS0 spectra in the FAST mode, background subtraction was not
done, area discrimination was used to select the source photons, and
hot and flickering pixels were not removed whose effect is relatively
small in this mode. Due to these effects, the spectral parameters obtained
from analysis of the SIS0 data are inaccurate. However, the pulse phase
dependence of the spectral components derived from the SIS0 data
are still reliable and agrees well with the results obtained from the
GIS data.

Though the pulse phase variation of the black-body flux of LMC~X-4 
is consistent with a sinusoidal nature, a very low pulse fraction in this
source and relatively large error bars makes it far from conclusive.
Since the pulse profile of LMC~X-4 in higher energy band shows several
narrow features, a coarse phase bin in LMC~X-4 will not show any
variation in the high energy band. We, therefore, only show the flux of
the black-body component obtained from four equal phase bins in the
bottom panel of Figure 10.
Since the statistics is poor in LMC~X-4, a comparison similar to
what is presented in Figure 6 for SMC~X-1, could not
be made.

\section{Discussion}

The pulse phase averaged spectra of these two accretion powered bright
X-ray pulsars obtained with ASCA clearly show the presence of a soft
component, which is in excess of the hard power-law component extended
to lower energies.
The present
work is a confirmation of the presence of soft excess in LMC~X-4 and
SMC~X-1.
Pulse phase resolved spectroscopy, presented here, clearly show that
in SMC X-1, the soft
component when fitted by a black-body (or thermal bremsstrahlung) model,
exhibits a sinusoidal pulse shape regardless of the complex, sharp profile
in hard energy band. The same may also be true for LMC X-4.
A dissimilar pulse profile between the low and
high energy bands is an indication that the soft component may have a
different origin of emission, or at least the geometry of emission is
different in different energy ranges.
In LMC X-4, La Barbera et al. (2001) also fitted the soft excess as
seed photon emission from accretion disk at magnetoshperic radius
Comptonized by moderately hot electrons. We have verified that this
model can fit the soft excess seen in the ASCA spectra of both SMC X-1
and LMC X-4 with electron temperature of $\sim$ 1.0 keV and optical
depth of 8--10. If the soft photons are produced from the vicinity
of the Alfven radius, a modulation of the soft excess at the pulse
period is difficult to explain.

Among the other X-ray pulsars known to have soft excess, in which pulse
phase resolved spectroscopy has also been done, RX~J0059.2--7138 (Kohno
et al. 2000) and X~Per (Coburn et al. 2001) do not show any
modulation in the soft component, XTE~0111.2--7317 shows a pulse modulation
of soft excess which is approximately in phase with the power-law
component (Yokogawa et al. 2000b), and Her~X-1 shows a pulse modulation
of soft excess which has a different phase of pulse peak compared to the
power-law component (Endo et al. 2000).
The luminosity in the soft X-ray excess in SMC~X-1 and LMC~X-4
are 8.4$\times$10$^{36}$ and 7.7$\times$10$^{36}$ erg s$^{-1}$
respectively, a factor of 7--8 larger compared to that of Her~X-1.
In Her~X-1, the soft excess, which is modeled as a
black-body type emission, probably comes from a part of the X-ray
irradiated inner accretion disk (Endo et al. 2000). In addition to the soft
component, there is also a broad low energy emission line with a pulse
shape and phase identical with that of the black-body emission. The total
amount of X-ray emission in the soft excess is $1.1 \times 10^{36}$ erg
s$^{-1}$, which is 10\% of the total emission in the ASCA energy band. In
comparison with Her X-1, the ratio of the black-body flux to the total
flux, is somewhat smaller in SMC~X-1 (3.6\%) and LMC~X-4 (6.4\%). In many
ways, SMC~X-1 and LMC~X-4 are quite similar to Her~X-1. It is interesting
to investigate whether the presence of the soft component and its pulsating
nature can be explained in a manner similar to Her~X-1.

The soft excess
can be a part of the X-ray emission from regions close to the neutron
star reprocessed by optically thick surrounding matter in the inner
accretion disk. With an assumption that the phase averaged pulsar emission
is isotropic, the reprocessing region holds a solid angle that is equal
to the fraction of X-ray energy reprocessed ($\sim$5\% in SMC~X-1 and
LMC~X-4). From energy conservation, the radius at which the reprocessing
region is situated ($r_{\rm BB}$), presumably the inner accretion disk or
Alfven radius,
can be calculated from the expression
$$
{{L_{\rm X}}\over {4\pi r_{\rm BB}^2}} = \sigma T^4,
$$
where $L_X$ is the
total X-ray luminosity and $T$ is the temperature of the black-body emission.
For SMC~X-1 ($L_{\rm X} = 2.4 \times 10^{38}$ erg s$^{-1}$ ) and LMC~X-4
($L_{\rm X} = 1.2 \times 10^{38}$ erg s$^{-1}$), $r_{\rm BB}$ is calculated
to be $12 \times 10^7$~cm and $9.4 \times 10^7$~cm respectively.

Here, we make a comparison of Alfven radius ($r_{\rm A}$) in these two
pulsars with the same in Her~X-1. For a given mass and radius of the
neutron star, the inner
radius scales as $r_{\rm A} \propto B^{4 \over 7}L_{\rm X}^{-{2 \over 7}}$. In
these two pulsars, $L_{\rm X}$ is larger by about one
order of magnitude than that in Her X-1.
The magnetic field strength ($B$) in SMC~X-1 is unknown, but the spectral
cutoff energy indicate that $B$ is smaller than that in Her~X-1
(Makishima et al. 1999). In LMC~X-4, a spectral cutoff at 16 keV
and a recently discovered cyclotron absorption feature at $\sim$100 keV
with the Beppo-SAX (La Barbera et al. 2001) are in disagreement with the
correlation observed in other accretion powered X-ray pulsars.
However, considering $L_{\rm X}$ and $B$ (corresponding to the Beppo-SAX
measurement of the cyclotron absorption feature),
the radius of the inner accretion disk is expected to be smaller in these
two pulsars compared to Her~X-1. On the contrary, in Her X-1, $r_{\rm BB}$
is found to be $3.6 \times 10^7$ cm (Endo et al. 2000), somewhat
smaller than $r_{\rm BB}$ estimated above for SMC~X-1 and LMC~X-4.
This is one reason we suspect that a black-body type emission cannot
satisfactorily explain the soft excess in these two bright X-ray pulsars.

However, absence of convincing direct measurement of the magnetic
field strength in these two pulsars result in some ambiguity in the
Alfven radius.
We also note that X-ray irradiated inner accretion disk can be
highly ionised with color temperature higher than the effective
temperature. Presence of weak undetected emission lines may also cause
the energy spectrum to deviate from a perfect blackbody, and the
estimated parameters for the black-body emission region may include
large systematic errors. Considering the above factors, we feel that
it may still be possible to interpret that the Alfven radius is the
emission site of the black-body type soft excess.

Regarding Model II for the spectrum, a large emission measure
($5 \times 10^{61}$ and $7\times10^{60}$ cm$^{-3}$ for SMC~X-1 and
LMC~X-4 respectively) required for the soft
excess indicates that with a thin thermal condition ($\tau \leq 1.0$),
the emission region has to be extremely large
($\ga 5 \times 10^{12}~{\rm cm}$ and $\ga 8\times10^{11}~{\rm cm}$ for SMC~X-1
and LMC~X-4 respectively). Since the light travel time across such a
region is much larger than the pulse periods in consideration,
a pulsating soft component can not be produced by such a plasma.
Moreover, even at extremely high densities, the cooling time scale
of the plasma will be larger than the pulse periods of these pulsars.
A pulsating nature of the soft component, therefore, cannot be
explained in the thermal bremsstrahlung model also.

The two alternate models for the pulsar spectrum that we have tried here give
somewhat satisfactory fit to the observed spectra. An additional soft
power-law or an inversely broken shape of the power-law may also arise
naturally when the entire emission is from the accretion column. The
distribution of temperature and matter density in the accretion
column and transparency of radiation of different energy to the magnetic
field structure determines the pulse profile and the energy spectrum.
The soft component probably arises from upper part of the accretion column
while the hard component is from the lower part. The different pulse
shape and pulse phases of the soft and hard spectral components in Her
X-1 have been compared with the model emission from an accretion mound
on the magnetic pole of a neutron star (Burnard, Arons, and Klein 1991).
Different types of
beaming at different energy bands and at different heights of the
accretion column can give rise to the observed energy dependence of the
pulse profiles. The high, near-Eddington luminosity of SMC~X-1 and LMC~X-4,
most of which is in the hard power-law component can be explained in a fan
beam type emission
pattern, in which the X-ray emission is perpendicular to the matter flow
in the accretion columns. Observations of these type of bright pulsars in
future, with instruments having better energy resolution and sensitivity
in the lower energy band may help us to find an exact model of the pulsar
spectrum.

\section{Conclusions}

In the present work we have found that these two high luminosity pulsars
show pulsations in the entire energy band of 0.5--10.0 keV, with strong
energy dependence in the pulse shape. The pulse phase averaged energy
spectra definitely show the presence of soft excess in both the sources.
If the soft excess is modeled as a separate black-body or thermal
bremsstrahlung type component, pulse phase resolved spectroscopy shows
that the soft component has a pulsating nature in SMC X-1, which may also
be true for LMC X-4.
We have found that the pulse profile of the soft component
is nearly sinusoidal, significantly different from the sharp, complex profile
of the hard power-law component. Due to very high
luminosity of these sources, the size of the emission zone required for
the soft component is large (radius $\sim$300--400 km) and we find that
it is difficult to explain the pulsations detected at low energies. We
have found that alternate models like inversely broken power-law or two
different power-law components can also be used to describe the spectra.

\begin{acknowledgements}
We thank R. Sunyaev and A. R. Rao for some useful discussions.
This research has made use of data obtained through the High Energy
Astrophysics Science Archive Research Center Online Service, provided by the
NASA/Goddard Space Flight Center. B. Paul was supported by the Japan Society
for the Promotion of Science through a fellowship during which most of this
work was done.
FN acknowledges the support by the Japan Science and Technology
Corporation under the ACT-JST project.
\end{acknowledgements}

\clearpage

\begin{deluxetable}{lcccc}
\footnotesize
\tablenum{1}
\tablecaption{The spectral parameters of SMC~X-1}
\tablewidth{0pt}
\tablehead{
\colhead{Parameter}&\colhead{Model I}&\colhead{Model II}&\colhead{Model III}&\colhead{Model IV}\nl}
\startdata                                                                      
N$_{\rm H}$\tablenotemark{a} & 2.16 $\pm$ 0.32 & 3.52 $\pm$ 0.31 & 2.10 $\pm$ 0.22 & 5.3 $\pm$ 0.5  \nl
Photon index ($\Gamma_1$) & 0.91 $\pm$ 0.03 & 0.93 $\pm$ 0.04 & 0.95 $\pm$ 0.02 & 0.87 $\pm$ 0.04  \nl
Power-law norm($I_{\rm PL}$)\tablenotemark{b}
                               & 2.98 $\pm$ 0.12 & 3.13 $\pm$ 0.16 & 5.18 $\pm$ 0.22 & 2.9 $\pm$ 0.2  \nl
Break energy (keV)             & --  &  --  &  1.68 $\pm$ 0.04  &  --  \nl
Photon index ($\Gamma_2$)      & --  &  --  &  1.90 $\pm$ 0.10  &  4.85 $\pm$ 0.30  \nl
Power-law norm($I_{\rm PL_{\rm S}}$)\tablenotemark{b}
                               & --  &  --  &  --  & 8.3 $\pm$ 1.2 \nl
BB/BR temp (keV)               & 0.179$\pm$ 0.009 & 0.33 $\pm$ 0.03 &  --  &  --  \nl
BB/BR norm\tablenotemark{c}    & 6750$^{+2000}_{-2500}$ & 0.46 $\pm$ 0.20 &  --                 &  --           \nl
Cutoff Energy ($E_{\rm C}$) keV& 5.5$^{+1.4}_{-0.5}$ & 6.3$^{+0.9}_{-2.0}$ & 5.5$^{+2.8}_{-0.7}$   & 5.7 $\pm$ 1.1       \nl
E-folding energy ($E_{\rm F}$) keV& 35$^{+23}_{-20}$ & 25$^{+25}_{-10}$ & 55 $\pm$  40& 25 $\pm$ 15 \nl
Fe Line centre energy (keV)    & 6.32$\pm$ 0.16      & 6.31$^{+0.17}_{-0.22}$      & 6.32 (fixed)        & 6.41$^{+0.17}_{-0.13}$\nl
Fe Line width (keV)            & 0.5 $\pm$ 0.2       & 0.5$^{+0.3}_{-0.2}$       & 0.50 (fixed)      & 0.42 $\pm$ 0.22 \nl
Fe Line norm\tablenotemark{d}  & 8.3$^{+6.5}_{-4.7}$       & 9.5 $\pm$ 6.5       & 9.5$^{+1.7}_{-2.5}$       & 6.1$^{+4.0}_{-3.5}$ \nl
Reduced $\chi^2$/dof           & 0.961/477           & 1.03/477           & 1.18/479           & 1.22/477      \nl

\tablenotetext{a}{$10^{21}$ atoms cm$^{-2}$}
\tablenotetext{b}{$10^{-2}$ photons cm$^{-2}$ s$^{-1}$ keV$^{-1}$ at 1 keV}
\tablenotetext{c}{
BB: $\left(\rm R_{\rm km}\over D_{\rm 10\,kpc}\right)^2$;~~~~~~~~~
BR: ${{3.2 \times 10^{-60}}\over{4\pi {D_{\rm 10\,kpc}}^2}}$
{$\int n_{\rm e} n_{\rm I} {\rm dV}$} cm$^{-3}$           }                              
\tablenotetext{d}{$10^{-4}$ photons cm$^{-2}$ s$^{-l}$ }

\enddata
 
\end{deluxetable}                                                               

\clearpage

\begin{deluxetable}{lcccc}
\footnotesize
\tablenum{2}
\tablecaption{Pulse phase resolved spectral parameters of SMC~X-1 for Model I}
\tablewidth{0pt}
\tablehead{
\colhead{Parameter}&\colhead{Main pulse}&\colhead{1st minima}&\colhead{Sub pulse}&\colhead{2nd minima}\nl}
\startdata                                                                      
N$_{\rm H}$\tablenotemark{a} & 6.0 $\pm$ 0.6 & 4.6 $\pm$ 1.2 & 7.9 $\pm$ 1.7 & 3.9 $\pm$ 1.9  \nl
Photon index ($\Gamma_1$) & 1.13 $\pm$ 0.02 & 1.09 $\pm$ 0.06 & 1.10 $\pm$ 0.07 & 1.05 $\pm$ 0.08  \nl
Power-law norm($I_{\rm PL}$)\tablenotemark{b}
                               & 2.05 $\pm$ 0.07 & 1.16 $\pm$ 0.11 & 1.55 $\pm$ 0.16 & 1.03 $\pm$ 0.13 \nl
BB temp (keV)               & 0.135 $\pm$ 0.004 & 0.142 $\pm$ 0.012 &  0.133 $\pm$ 0.009  &  0.149 $\pm$ 0.017  \nl
BB flux\tablenotemark{c}~(0.7-10.0 keV) &	1.14	& 0.81 & 0.84 & 0.67\nl
BB flux 	 &	5.0 & 5.9 & 4.9 & 5.1 \nl
(\% of total flux in 0.7-10.0 keV) \nl
Reduced $\chi^2$/dof           & 1.16/599           & 0.92/304           & 1.00/306           & 0.87/245      \nl

\tablenotetext{a}{$10^{21}$ atoms cm$^{-2}$}
\tablenotetext{b}{$10^{-2}$ photons cm$^{-2}$ s$^{-1}$ keV$^{-1}$ at 1 keV}
\tablenotetext{c}{$10^{-11}$ erg cm$^{-2}$ s$^{-l}$ }

\enddata
 
\end{deluxetable}                                                               

\clearpage

\begin{deluxetable}{lcccc}
\footnotesize
\tablenum{3}
\tablecaption{The spectral parameters of LMC~X-4}
\tablewidth{0pt}
\tablehead{
\colhead{Parameter}&\colhead{Model I}&\colhead{Model II}&\colhead{Model III}&\colhead{Model IV}\nl}
\startdata                                                                      
N$_{\rm H}$\tablenotemark{a} & 5.7$^{+3.8}_{-0.0}$   & 5.7$^{+3.2}_{-0.0}$   &
5.7$^{+1.3}_{-0.0}$   &   5.7$^{+5.5}_{-0.0}$   \nl
Photon index ($\Gamma_1$) & 0.69 $\pm$ 0.04       & 0.75 $\pm$ 0.05       & 0.73 $\pm$ 0.03       & 0.59 $\pm$ 0.04 \nl
Power-law norm($I_{\rm PL}$)\tablenotemark{b}
                               & 1.48 $\pm$ 0.07       & 1.41 $\pm$ 0.10       & 2.66 $\pm$ 0.06       & 1.07 $\pm$ 0.08 \nl
Break energy (keV)             &  --                   &  --                   & 1.75 $\pm$ 0.04 &  --             \nl
Photon index ($\Gamma_2$)      &  --                   &  --                   & 1.94 $\pm$ 0.07 & 2.9 $^{+0.7}_{-0.2}$  \nl
Power-law norm($I_{\rm PL_{\rm S}}$)\tablenotemark{b}
                               &  --                   &  --                   &  --                   &  1.3 $\pm$ 0.3  \nl
BB/BR temp (keV)               & 0.170$^{+0.12}_{-0.18}$        & 0.51 $\pm$ 0.04       &  --                   &  --             \nl
BB/BR norm\tablenotemark{c}    & 4660 $^{+4100}_{-1100}$        & 0.053
$^{+0.025}_{-0.007}$        &  --                   &  --             \nl
L1 Line centre energy (keV)    & 0.96 $\pm$ 0.04       & 0.93 $\pm$ 0.03       & 0.95 $\pm$ 0.03       & 0.90 $\pm$ 0.03 \nl
L1 Line norm\tablenotemark{d}  & 19 $\pm$ 4        & 35 $\pm$ 9        & 21 $\pm$ 5        & 56 $\pm$ 13  \nl
L2 Line centre energy (keV)   & 1.93 $\pm$ 0.04       & 1.94 $\pm$ 0.05       & 1.88 $\pm$ 0.03       & 1.93 $\pm$ 0.05 \nl
L2 Line norm\tablenotemark{d} & 3.0 $\pm$ 0.7         & 1.6 $\pm$ 0.6         & 4.0 $\pm$ 0.8         & 1.1 $\pm$ 0.6   \nl
Fe Line centre energy (keV)    & 6.47 $\pm$ 0.12       & 6.49 $\pm$ 0.11       & 6.56$^{+0.18}_{-0.12}$       & 6.44 $\pm$ 0.12 \nl
Fe Line width (keV)            & 0.52 $\pm$ 0.15         & 0.57 $\pm$ 0.15         & 0.68 $\pm$ 0.20       & 0.55 $\pm$ 0.15  \nl
Fe Line norm\tablenotemark{d}  & 6.5 $\pm$ 2.0   & 7.4 $\pm$ 2.1  & 10 $\pm$ 4 & 7.0 $\pm$ 2.0 \nl
Reduced $\chi^2$/dof           & 1.38/340             & 1.37/339             & 1.35/340             & 1.31/339       \nl

\tablenotetext{a}{$10^{20}$ atoms cm$^{-2}$}
\tablenotetext{b}{$10^{-2}$ photons cm$^{-2}$ s$^{-1}$ keV$^{-1}$ at 1 keV}
\tablenotetext{c}{
BB: $\left(\rm R_{\rm km}\over D_{\rm 10\,kpc}\right)^2$;~~~~~~~~~
BR: ${{3.2 \times 10^{-60}}\over{4\pi {D_{\rm 10\,kpc}}^2}}$
{$\int n_{\rm e} n_{\rm I} {\rm dV}$}  cm$^{-3}$          }                              
\tablenotetext{d}{$10^{-4}$ photons cm$^{-2}$ s$^{-1}$}

\enddata
 
\end{deluxetable}                                                               

\clearpage

\begin{figure}[t]
\plotone{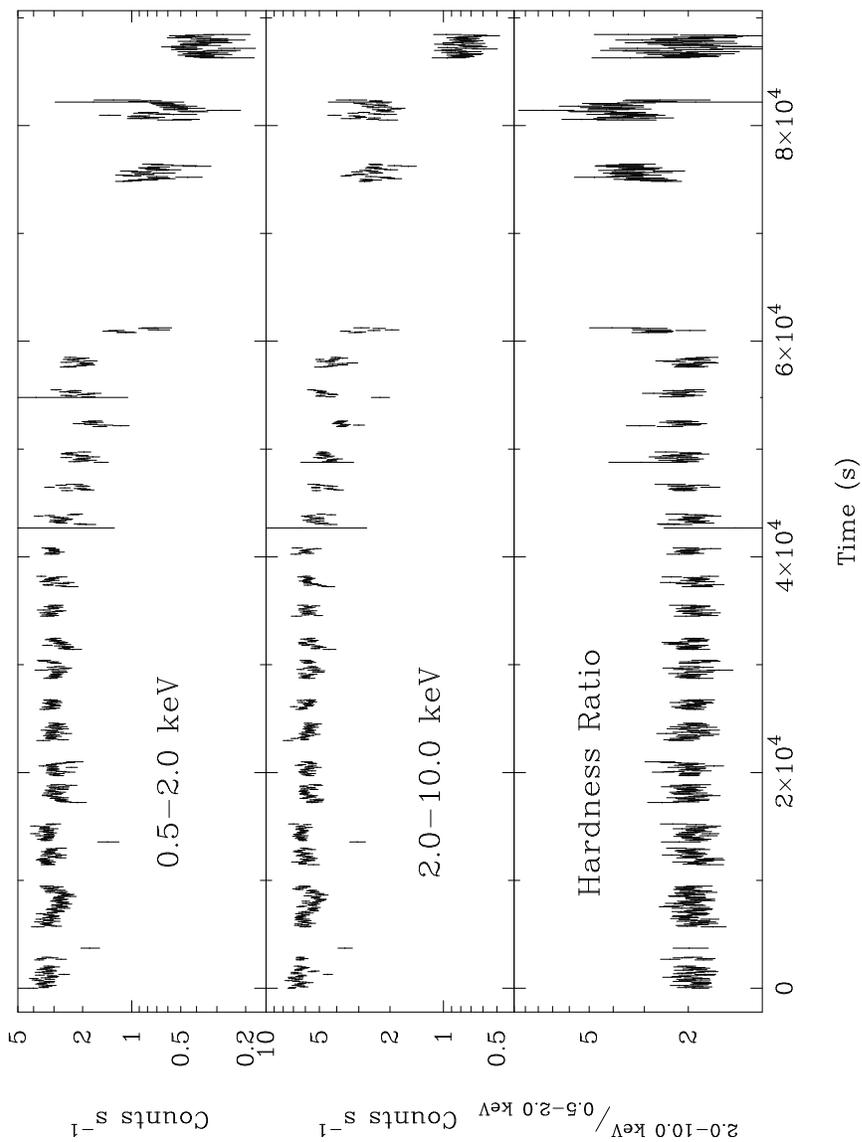}
\caption
{The light curves of SMC~X-1 in two energy bands of 0.5--2.0 and 2.0--10.0
keV and the hardness ratio obtained from the GIS data of the 1993
observation are shown here. Spectral changes can be noticed near the end
of the observation. For the spectral analysis, data from the first 60\,000 s
of this observation was used.}\label{fig1}
\end{figure}

\begin{figure}[t]
\plotone{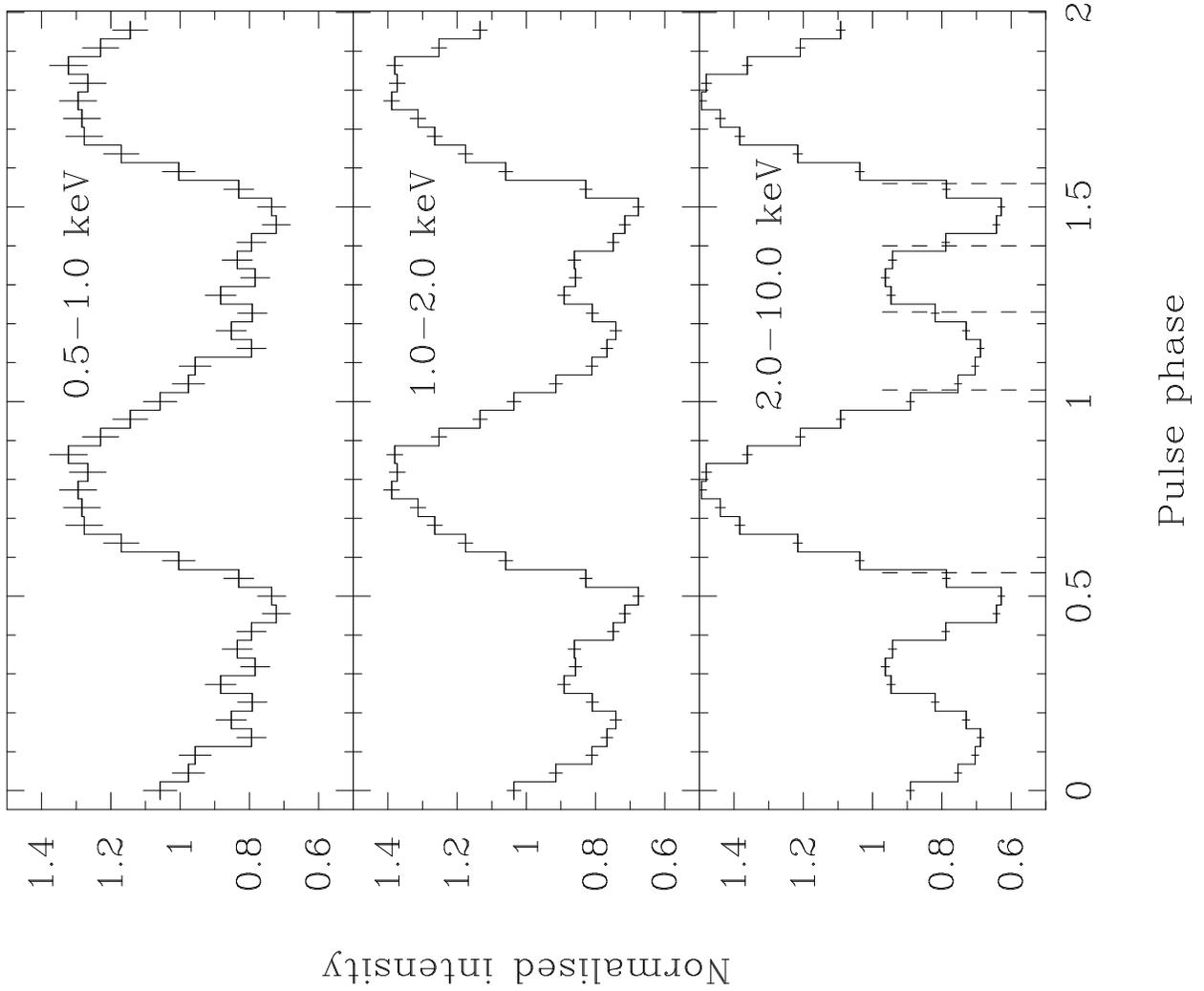}
\caption
{The normalised pulse profiles of SMC~X-1 in three energy bands obtained with
the GIS in the 1993 observation are plotted here. The main peak, sub peak
and the two minima regions chosen for coarse phase resolved spectroscopy are
marked in the bottom panel.}\label{fig2}
\end{figure}

\begin{figure}[t]
\epsscale{.8}
\plotone{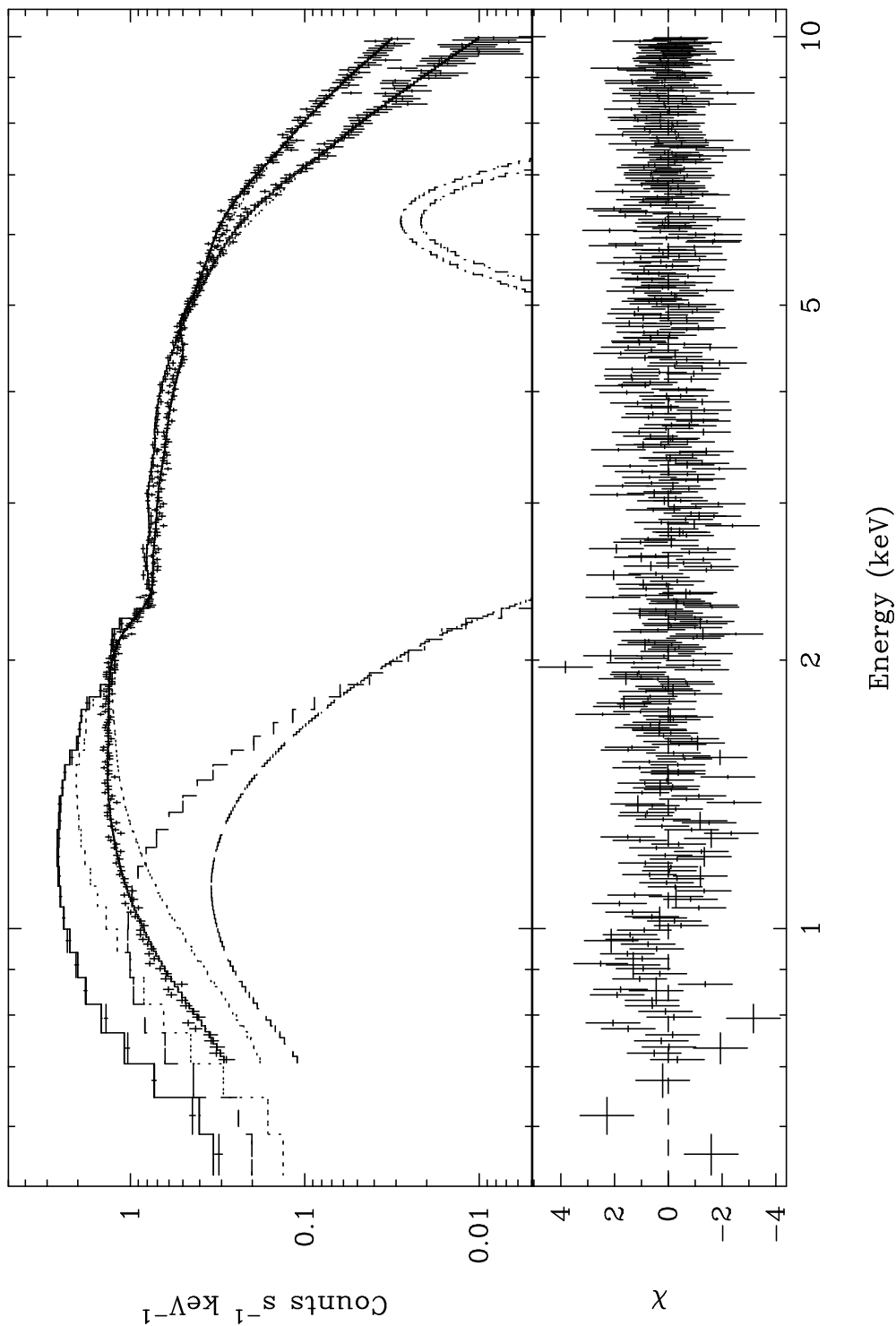}
\caption
{The energy spectra of SMC~X-1 obtained with the SIS and GIS are shown here
along with the best fit model comprising three components, an exponentially
cutoff power-law, a broad iron line emission, and a soft black-body emission.
The lower panel shows the contributions of the residuals to the $\chi^2$ for
each energy bin.}\label{fig3}
\end{figure}

\begin{figure}[t]
\plotone{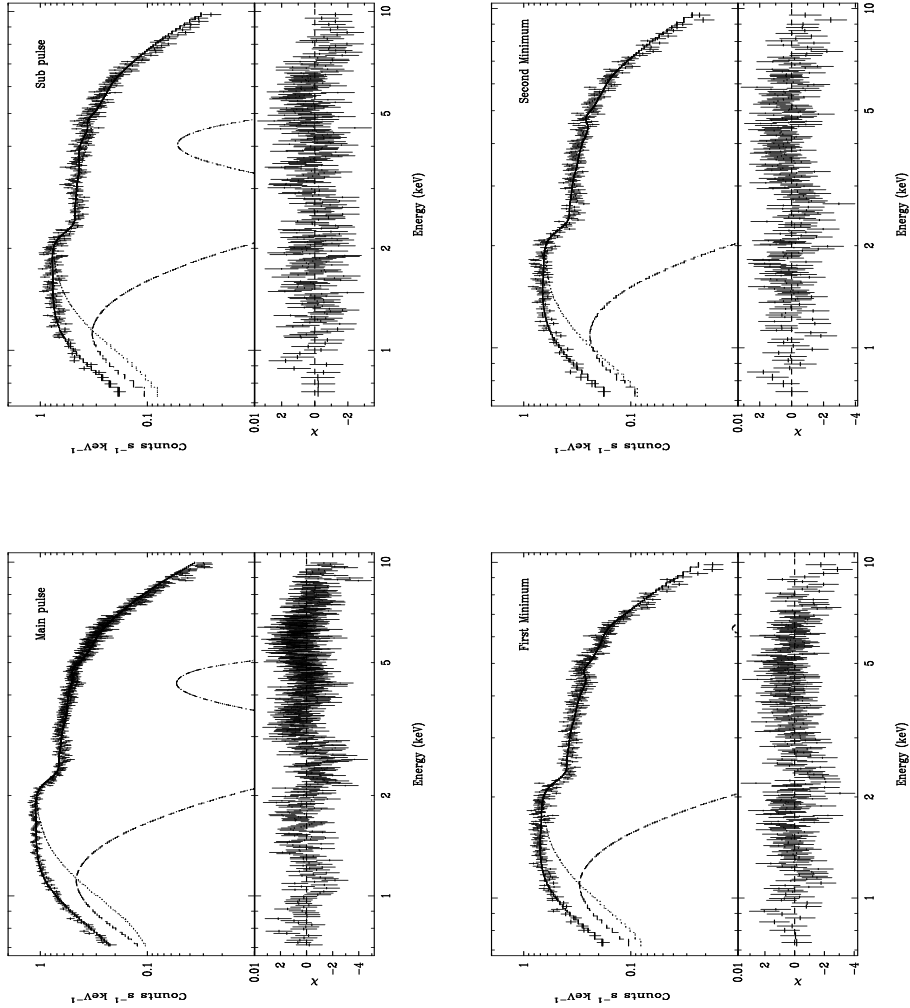}
\caption{Phase resolved spectra of SMC X-1 obtained with ASCA-GIS.}\label{fig4}
\end{figure}

\begin{figure}[t]
\plotone{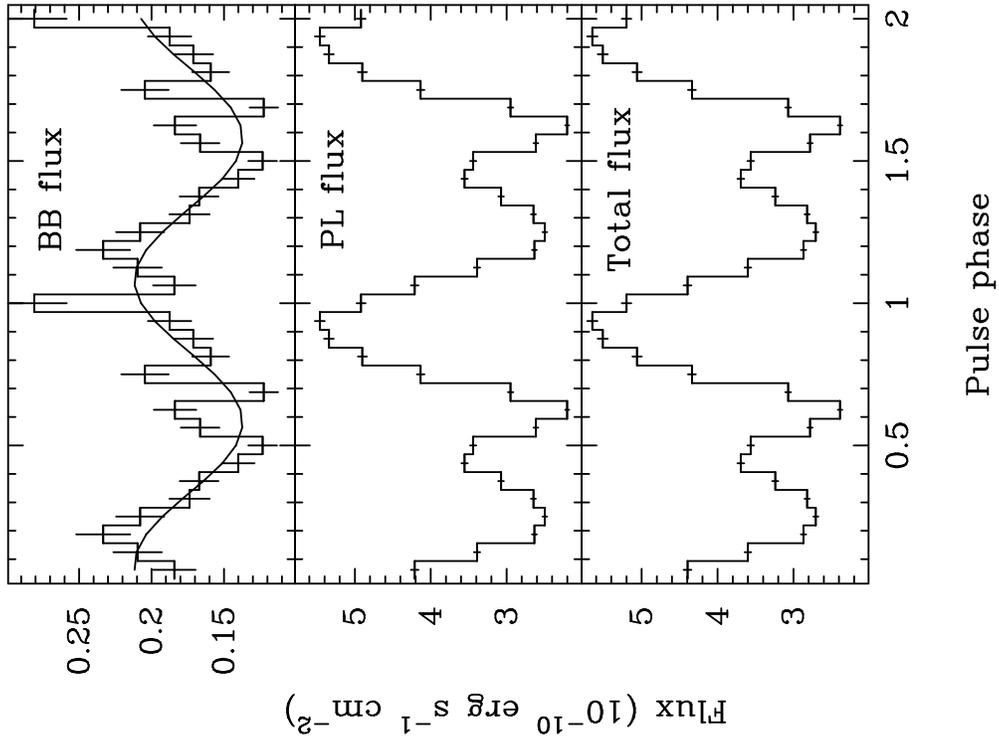}
\caption
{Modulation of the black-body flux, power-law flux, and total flux in the
0.5--10.0 keV band of SMC~X-1 obtained from pulse phase resolved analysis
of the GIS spectrum with Model I are shown here for two cycles. With
Model II, the bremsstrahlung component was found to have pulsations
similar to the black-body component shown here. The curve in the top
panel is the best fit sinusoidal, given here to indicate the almost
sinusoidal nature of the soft excess modulation.}\label{fig5}
\end{figure}

\begin{figure}[t]
\plotone{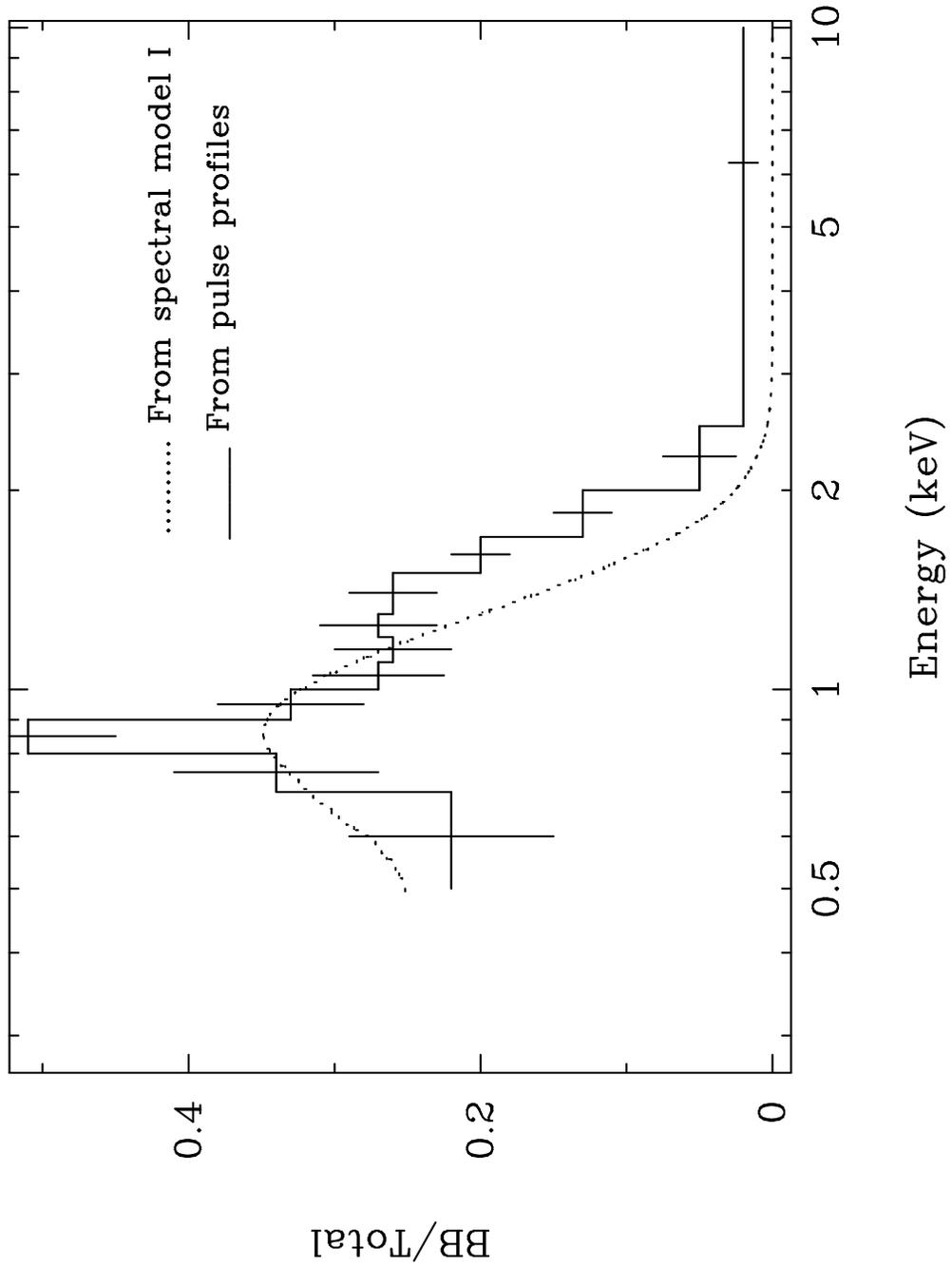}
\caption
{Ratio of the black-body to total flux from the spectral model is
compared with the fraction of sinusoidal (black-body) component
in the SMC~X-1 pulse profile at different energies}\label{fig6}
\end{figure}

\begin{figure}[t]
\plottwo{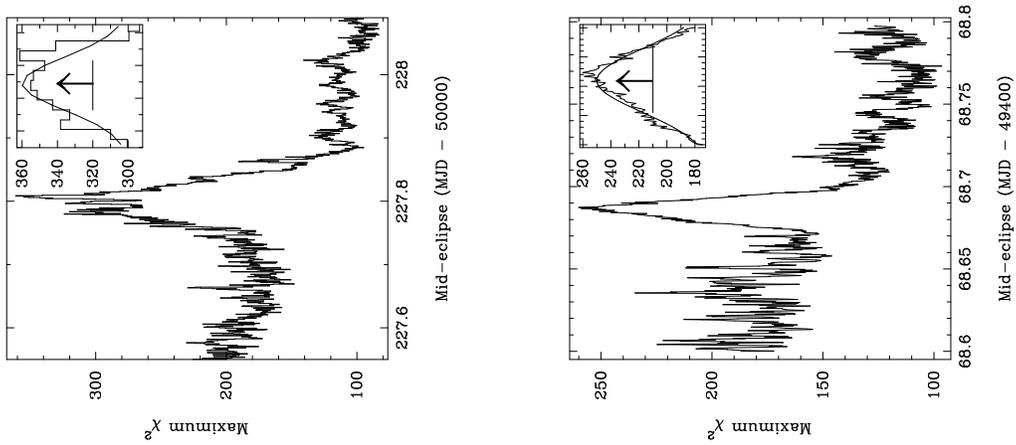}{f7b.ps}
\caption
{The maximum $\chi^2$ obtained from pulse folding technique is plotted
here against the trial mid-eclipse epochs for the 1994 and 1996 ASCA
observations of LMC~X-4. Expanded view of the region of interest is shown
in the inset of each panel. The center and width of the
best fitted gaussians are marked with arrow and horizonal line.}\label{fig7}
\end{figure}

\begin{figure}[t]
\plotone{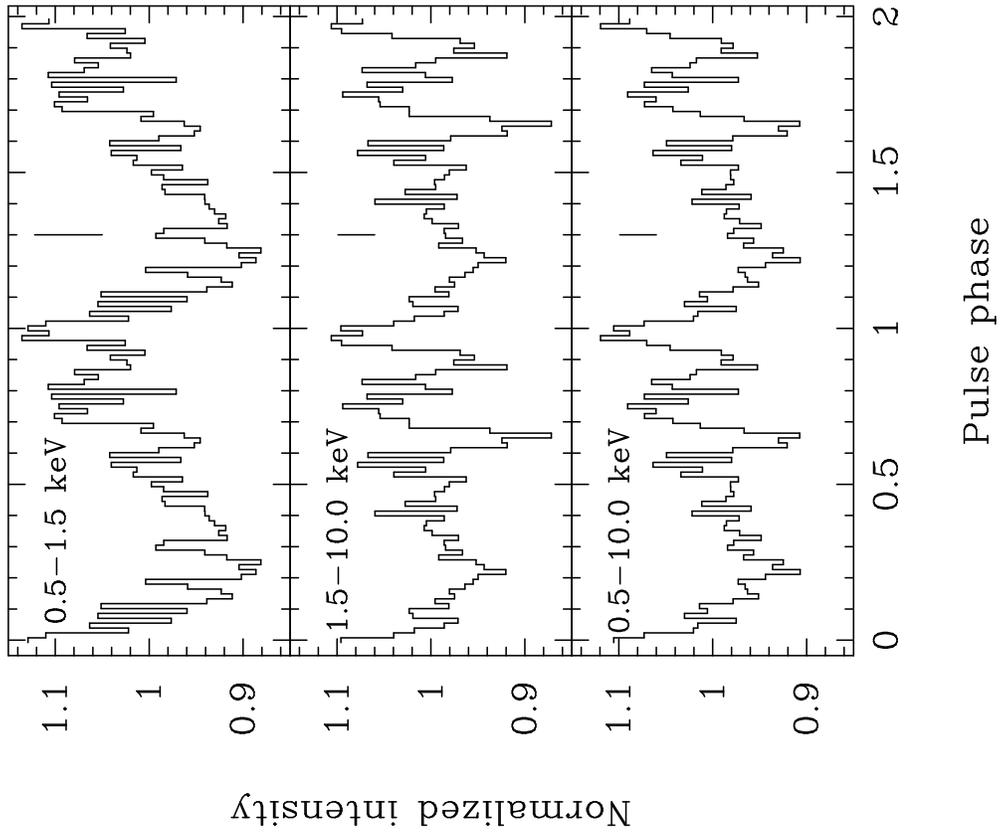}
\caption
{The normalised pulse profiles of LMC~X-4 in three energy bands obtained
with the ASCA-GIS are shown here. Typical error bars are given in each
panel.}\label{fig8}
\end{figure}

\begin{figure}[t]
\epsscale{.8}
\plotone{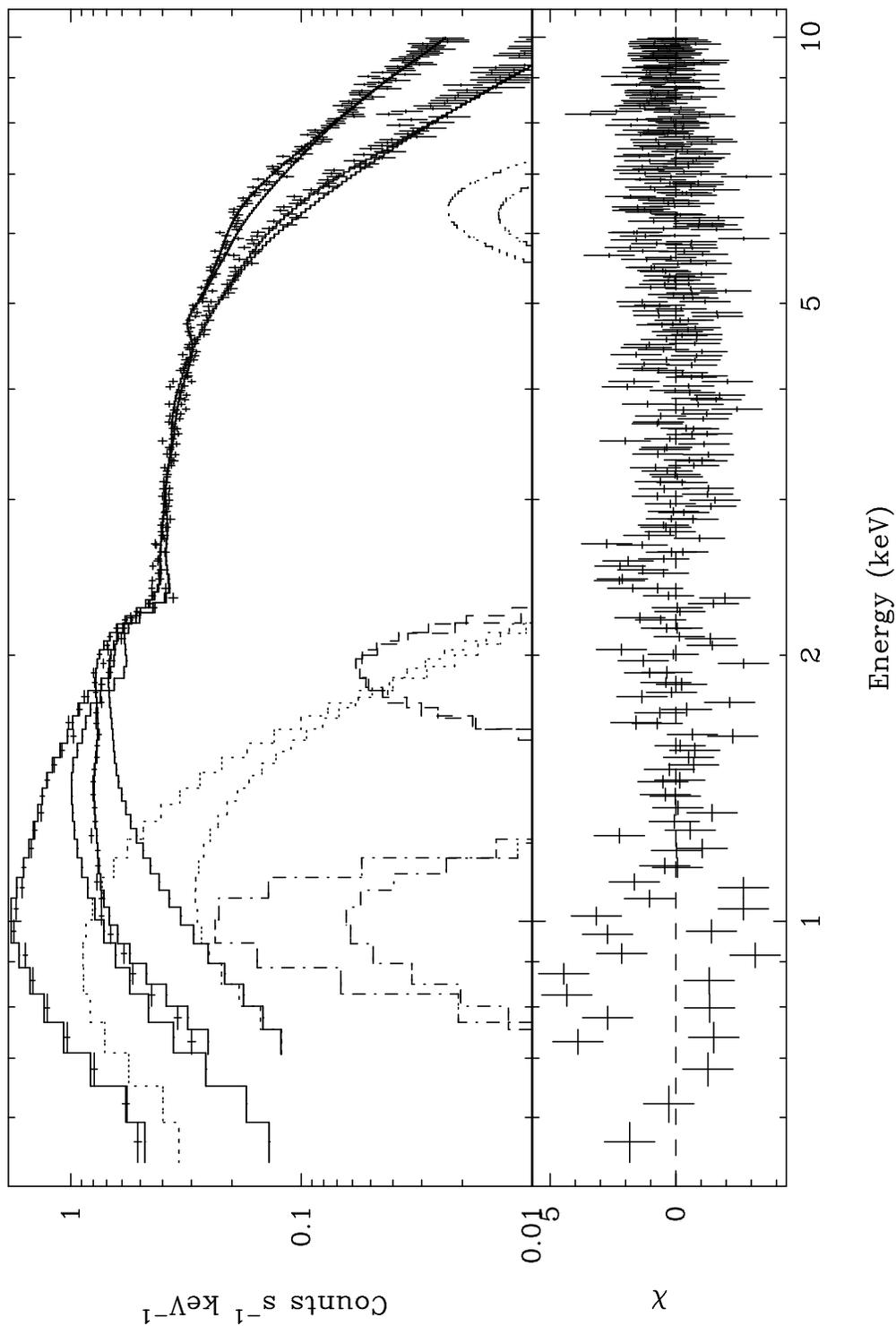}
\caption
{The energy spectra of LMC~X-4 obtained with the SIS1 and two GIS detectors
are shown here along with the best fit model comprising an absorbed power-law,
a broad iron line emission, a soft black-body emission and two low energy
emission lines. The lower panel shows the contributions of the residuals
to the $\chi^2$ for each energy bin.}\label{fig9}
\end{figure}

\begin{figure}[t]
\plotone{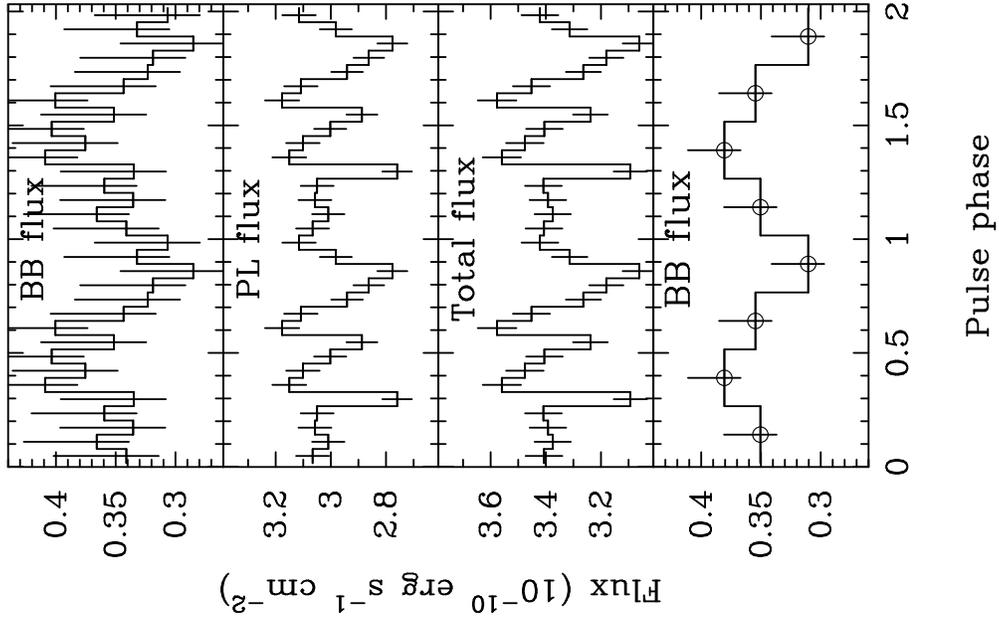}
\caption
{Modulation of the black-body flux, power-law flux, and total flux in the
0.5--10.0 keV band obtained from pulse phase resolved analysis of GIS
spectrum of LMC~X-4 with Model I are shown in the top three panels for
two cycles. With
Model II, the bremsstrahlung component was found to have pulsations
similar to the black-body component shown here.
Variation of the black body flux, obtained from a coarse phase binning
is shown in the bottom panel.
}\label{fig10}
\end{figure}

\end{document}